\documentstyle[manuscript,aps,version2]{revtex}
\draft
\begin{document}

\preprint{HEP-95-16}
\title
\bf Universality in Random Walk Models with Birth and Death
\endtitle
\author{Carl M. Bender}
\instit
Department of Physics, Washington University, St. Louis, MO 63130
\endinstit
\author{Stefan Boettcher}
\instit
Department of Physics, Brookhaven National Laboratory, Upton, NY
11973 \endinstit
\author{Peter N. Meisinger}
\instit
Department of Physics, Washington University, St. Louis, MO 63130
\endinstit
\medskip
\centerline{\today}
\abstract
Models of random walks are considered in which walkers are born at one location
and die at all other locations with uniform death rate. Steady-state
distributions of random walkers exhibit dimensionally dependent critical
behavior as a function of the birth rate. Exact analytical results for a
hyperspherical lattice yield a second-order phase transition with a nontrivial
critical exponent for all positive dimensions $D\neq 2,~4$. Numerical studies of
hypercubic and fractal lattices indicate that these exact results are universal.
Implications for the adsorption transition of polymers at curved interfaces are
discussed.
\endabstract


In this letter we study random walk models in which walkers are born at one
location and die everywhere else. We believe that our analytical results
for the critical behavior exhibited by {\sl spherically symmetric} random walks
are universal and therefore provide insights into the dynamics governing
dissipative systems that interact at an interface \cite{weiss}. As an
application, we discuss the adsorption transition of a polymer near
an attractive boundary \cite{Taka}.

Our spherically symmetric random walk model uses
a lattice that drastically simplifies the treatment of a dissipative system
interacting with a curved boundary. For example, we can model polymer growth
near a hyperspherical boundary in any dimension $D>0$ by a random walk on a
{\sl one-dimensional} lattice. Such a lattice faithfully represents the spatial
entropy of the system. While this lattice is not translationally
invariant, it is suitable for the study of spherically symmetric boundary-value
problems. To demonstrate the utility of this approach we compare numerical
results obtained from hypercubic and fractal lattices with our exact analytical
results and show that these results are universal.

A random walk on a lattice is described by a single probability function,
$C_{{\bf n},t;{\bf m}}$, the probability that a random walker who begins at
location ${\bf m}$ at $t=0$ will be at location ${\bf n}$ at time $t$. The
probability $C_{{\bf n},t;{\bf m}}$ satisfies a difference equation of generic
form:
\begin{eqnarray}
C_{{\bf n},t;{\bf m}} = \sum_{\{ \sigma \}_{\bf n}} P(\sigma\to
{\bf n}) C_{\sigma,t-1;{\bf m}},
\label{e1}
\end{eqnarray}
where $\{\sigma\}_{\bf n}$ is the set of locations $\sigma$ adjacent to the
location ${\bf n}$ and $P({\bf n}\to {\bf n}')$ represents the probability that
a walker at the location ${\bf n}$ will go to the location ${\bf n}'$ in one
step. The probability $C_{{\bf n},t;{\bf m}}$ obeys the initial condition
$C_{{\bf n},0;{\bf m}}=\delta_{{\bf n},{\bf m}}$. Local conservation of
probability is expressed by
$\sum_{\{ \sigma \}_{\bf n}} P({\bf n}\to\sigma)=1$ for all $\bf n$. 

In this letter we generalize the difference equation (\ref{e1}) to
allow for creation and annihilation of random walkers.
Specifically, we let random walkers give birth in one location,
which we label ${\bf 0}$, with birth rate $a$ and die in all other
locations with uniform death rate $z$. Walkers are created or
destroyed in a given location in proportion to the number of
walkers at that site, where $a$ and $z$ are the constants of
proportionality. Of course, $a$ acts as a birth rate if $a>1$; if
$a<1$, it is really a death rate. A similar interpretation applies to $z$.

Birth rates and death rates are properties of populations rather
than of single individuals. Thus, rather than solving an initial-value
problem for a single random walker, we study a large
population of random walkers. We represent such a population by a
distribution $G_{{\bf n},t}$, which denotes the number of random
walkers at location ${\bf n}$ at time $t$. The distribution
$G_{{\bf n},t}$ satisfies the same recursion relation as $C_{{\bf
n},t;{\bf m}}$ except for factors of $a$ and $z$: 
\begin{eqnarray}
G_{{\bf n},t}=\cases{
z\sum_{\{\sigma \}_{\bf n}} P(\sigma\to {\bf n}) G_{\sigma,t-1}&
$({\bf n}\not \in \{\sigma\}_{\bf 0}),$\cr 
\noalign{\smallskip}
a P({\bf 0}\to{\bf n}) G_{{\bf 0},t-1}+
z\sum_{\{\sigma\neq{\bf 0}\}_{\bf n}} P(\sigma\to {\bf n})
G_{\sigma,t-1}& $({\bf n}\in\{\sigma\}_{\bf 0}).$\cr }
\label{e4}
\end{eqnarray}
Note that $G_{{\bf n},t}\geq 0$ for all ${\bf n}$ and $t$. We are
interested in the asymptotic behavior of $G_{{\bf n},t}$ as
$t\to\infty$. The specific choice of initial distribution $G_{{\bf
n},0}$ is unimportant, aside from the requirement that it be
normalizable, because a random walk is a diffusive (dissipative)
process and details of $G_{{\bf n},0}$ are irretrievably lost as
time evolves; all initial distributions lead to the same large-time
behavior. We are interested in steady-state solutions
of Eq.~(\ref{e4}); the existence of such solutions imposes a
relationship between the birth and death rates.

To describe a dissipative system interacting with a hyperspherical boundary we
formulate our random walk model on a spherically symmetric lattice in
$D$-dimensional space \cite{BeBoM}. We consider the boundary to lie at the
origin. (The general case of a hyperspherical boundary of arbitrary radius is
considered in Ref.~\cite{rw3}.) We take location $n$ to be the volume bounded by
two concentric $D$-dimensional hyperspherical surfaces of radii $R_{n-1}$ and
$R_n$. If the random walker is at location $n$ at time $t$, then at time $t+1$
the walker must move outward to location $n+1$ with probability $P_{\rm out}(n)$
or inward to location $n-1$ with probability $P_{\rm in}(n)$, where $P_{\rm out}
(n)+P_{\rm in}(n)=1$ so that probability is locally conserved. Because a walker
at the origin ($n=1$) must move outward, we take $P_{\rm out}
(1)=1$ and $P_{\rm in}(1)=0$. Then, the equation for a $D$-dimensional random
walk in Eq.~(\ref{e4}) reduces to a one-dimensional recursion relation:
\begin{eqnarray}
G_{n,t}=\cases{z P_{\rm in}(n+1) G_{n+1,t-1}+z P_{\rm out}(n-1)G_{n-1,t-1}&
$(n\geq 3),$\cr 
\noalign{\smallskip}
z P_{\rm in}(3) G_{3,t-1}+a G_{n-1,t-1}& $(n=2),$\cr 
\noalign{\smallskip}
z P_{\rm in}(2) G_{2,t-1}& $(n=1).$\cr}
\label{e15}
\end{eqnarray}

In Ref.~\cite{BeBoM} we take the probabilities of moving out or in to be in
proportion to the hyperspherical surface areas that can be traversed on each
move. Let $S_D(R)=2\pi^{D/2} R^{D-1}/\Gamma(D/2)$ represent the surface area
of a $D$-dimensional hypersphere. Then, for $n>1$,
\begin{eqnarray}
P_{\rm out}(n)={{S_D(R_n)}\over{S_D(R_n)+S_D(R_{n-1})}}\quad {\rm and}\quad
P_{\rm in}(n)={{S_D(R_{n-1})}\over{S_D(R_n)+S_D(R_{n-1})}}. 
\label{e20}
\end{eqnarray}

{}For dimensions other than $D=1$ and $D=2$, when we take $R_n=n$, the
difference equation in (\ref{e15}) cannot be solved in closed form \cite{BoMo}.
Thus, we propose that the probabilities in Eqs.~(\ref{e20}) be replaced by
bilinear functions of $n$, which are uniformly good approximations to
$P_{\rm out}(n)$ and $P_{\rm in}(n)$ in the range $D>0$ when $R_n=n$
\cite{BeCoMe}:
\begin{eqnarray}
P_{\rm out}(n)={n+D-2\over 2n+D-3}\quad {\rm and}\quad P_{\rm in}(n)={n-1\over
2n+D-3}.
\label{e22}
\end{eqnarray}
In Ref.~\cite{rw2} it was shown that this crucial simplification in $P_{\rm in}$
and $P_{\rm out}$ preserves the configurational entropy. Now, Eq.~(\ref{e15})
can be solved in closed form for all $D>0$:
\begin{eqnarray}
G_{n,t}=z^t~{\Gamma\left({D+1\over 2}\right)\Gamma(n+D-2)\over 2^{n-2} \Gamma(D)
\Gamma\left(n+{D-3\over 2}\right)} \oint_C {dy\over 2\pi i y}~y^{n-t-1}{\left(
{z\over a}\right)^{\delta_{1,n}}{}_2F_1\left({n\over 2},{n+1\over 2};n+{D-1\over
2}; y^2\right)\over 1+\left({z\over a}-1\right){}_2F_1\left({1\over 2},1;{D+1
\over 2}; y^2\right)}.
\label{e23}
\end{eqnarray}
{}From the solution in Eq.~(\ref{e23}) we can obtain closed-form expressions for
spatial and temporal moments of the random walk.

To study critical behavior in the random walk models in Eq.~(\ref{e4}) or
Eq.~(\ref{e15}) we take $F$, the fraction of walkers at the boundary ${\bf 0}$
(the location where random walkers are born), as our order parameter. For the
case of polymer adsorption, this observable corresponds to the fraction of the
polymer adsorbed on the boundary. To construct the fraction $F$ we let $N_t=
\sum_{\bf n}G_{{\bf n},t}$ be the total number of random walkers at time $t$. We
restrict our attention to initial distributions for which $N_0$ is finite so
that $N_t$ is finite for all $t$. Then, $F_t=G_{{\bf 0},t}/N_t$ represents the
fraction of all random walkers at location ${\bf 0}$ at time $t$.

We find that, independent of the choice of $G_{{\bf n},0}$, the asymptotic
behaviors of $N_t$ and $F_t$ are determined by the values of $a$ and $z$. 
Specifically, we obtain the generic (lattice-independent) result that the
positive quadrant of the $(a,z)$ plane is partitioned into four distinct regions
by two boundary curves as shown in Fig.~\ref{f1}. One boundary curve, labeled
$B_1$ in Fig.~\ref{f1}, is a straight line passing through the origin. To the
left of $B_1$, $F_t$ vanishes as $t\to\infty$; to the right of $B_1$, $F_t$
approaches a positive finite value as $t\to\infty$. When the lattice dimension
$D\leq 2$ the equation for the boundary line $B_1$ is $z=a$; as $D$ increases
beyond $2$ the boundary line remains straight but the slope of $B_1$ begins to
decrease with increasing $D$. The transition at $D=2$ is a reflection of Polya's
theorem \cite{Polya}, which states that when $D>2$ the probability of an 
individual random walker returning to a location is less than unity.

The second boundary curve shown in Fig.~\ref{f1} is labeled $B_2$. The first
part of $B_2$ is a straight-line segment, $z=1$, extending from the $z$ axis to
the boundary line $B_1$. This line segment connects to the second part of $B_2$,
which is a curve that approaches $z=0$ as $a\to\infty$. The equation for this
second part depends on $D$. [For a $D=1$ lattice, where the probabilities of
moving to the left or right are both $1\over 2$, this curve is given by $z=2a/
(a^2+1)~(a\geq 1)$.] Above $B_2$, $N_t\to\infty$ as $t\to\infty$; below $B_2$,
$N_t\to 0$ as $t\to\infty$. On $B_2$ the total number of walkers approaches a
finite number $N(a)$ as $t\to\infty$. On the curved portion of $B_2$, $N(a)>0$;
on the straight-line portion of $B_2$, $N(a)>0$ for $D>2$, while $N(a)=0$ for
$D\leq 2$. This transition at $D=2$ is yet another manifestation of Polya's
theorem.

As we cross the boundary line $B_1$, the limiting value of the function $F_t$ as
$t\to\infty$ is continuous. We are particularly interested in crossing from one
side of $B_1$ to the other along the boundary curve $B_2$ that divides the upper
region, where $N_t\to\infty$, and the lower region, where $N_t\to 0$ as $t\to
\infty$. The interpretation of $\lim_{t\to\infty}N_t$ being finite and nonzero
is that the distribution $G_{{\bf n},t}$ approaches a steady-state, where there
is a balance between random walkers being created at location ${\bf 0}$ and
annihilated in all other regions.

We focus on $B_2$ because it is only on this curve that a steady state is
reached as $t\to\infty$. Along $B_2$, $F(a)=\lim_{t\to \infty} F_t$
undergoes a second-order phase transition at the critical point $(a_{\rm c},
z_{\rm c}=1)$, which is situated at the intersection of $B_1$ and $B_2$. On
$B_2$, $F(a)=0$ when $a<a_{\rm c}$ (even though the limiting value of
$N_t$ may be $0$), and both $N(a)$ and $F(a)$ are {\sl finite} positive numbers
when $a>a_{\rm c}$. The curved portion of $B_2$ is the locus of all points in
the positive quadrant of the $(a,z)$ plane for which both $N(a)$ and $F(a)$ are
finite and nonzero.

Many of the qualitative features of Fig.~\ref{f1} can be derived directly from
an analysis of Eq.~(\ref{e4}). To determine the boundary line $B_1$ we introduce
a change of variable: $G_{{\bf n},t}=z^tH_{{\bf n},t}$. The distribution $H_{{
\bf n},t}$ satisfies the recursion relation for a random walk with birth rate
$a/z$ at location ${\bf 0}$ and no births or deaths at any other location. Let
$\Pi_{\bf 0}$ denote the probability that a random walker at location ${\bf 0}$
will ever return to location ${\bf 0}$. Suppose a random walk with birth rate
$a/z$ and death rate $1$ begins at $t=0$. Of the $H_{{\bf 0},0}$ walkers who
begin at location ${\bf 0}$, only a fraction $\Pi_{\bf 0}$ of them will return
to location ${\bf 0}$ to give birth to new walkers. Of these new walkers, again
only $\Pi_{\bf 0}$ of them will return to location ${\bf 0}$ to give birth
again, and so on. Hence, to find the total number of walkers who are ever born
we must sum a geometric series whose geometric ratio is the quantity $a\Pi_{\bf
0}/z$. If $a\Pi_{\bf 0}/z<1$, the geometric series converges and the total
number of random walkers ever born is finite. As time $t$ increases, the walkers
diffuse away from location ${\bf 0}$. Thus, as $t\to\infty$, the ratio
\begin{eqnarray}
{}F_t={G_{{\bf 0},t}\over\sum_{\bf n}G_{{\bf n},t}}={H_{{\bf
0},t}\over \sum_{\bf n} H_{{\bf n},t}}
\nonumber
\end{eqnarray}
vanishes. On the other hand, if $a\Pi_{\bf 0}/z>1$, both $H_{{\bf 0},t}$ and
$\sum_{\bf n} H_{{\bf n},t}$ diverge at the same rate, and the ratio $F_t$
approaches a nonzero limit (that lies between 0 and 1) as $t\to\infty$.

The transition between $F_t\to 0$ and $F_t\to finite~limit$ occurs on the line
$z=a\Pi_{\bf 0}$. This is the equation of the boundary line $B_1$. Polya's
theorem states that for any random walk $\Pi_{\bf 0}=1$ when $D\leq 2$ and
$\Pi_{\bf 0}<1$ when $D>2$. This theorem explains the transition in the slope of
the line $B_1$ at $D=2$ \cite{Polya}. In particular, since $z_{\rm c}=1$ 
for all $D>0$, we find $a_{\rm c}=1/\Pi_{\bf 0}$.

The shape of the curved part of $B_2$ depends on the choice of lattice; it is
not universal. However, the straight-line portion of $B_2$ is universal and is
easy to understand. Points $(a,z)$ such that $a<a_{\rm c}$ and $z$ near $1$ lie
to the left of $B_1$. Thus, $F_t$ becomes vanishingly small as $t\to\infty$.
Hence, the effect of the birth rate $a$ on the total number of walkers is
negligible. The growth or decay of the total number of walkers depends only on
$z$; if $z<1$ then $N_t\to 0$ as $t\to\infty$, and if $z>1$ then $N_t\to\infty$
as $t\to\infty$.

On the straight-line portion of the curve $B_2$, where $a<a_{\rm c}$ and $z=1$,
the limiting value of $N_t$ depends on the dimension $D$. If $D\leq 2$ then $a_
{\rm c}=1$. Thus, on this portion of $B_2$ a fraction $1-a$ of random walkers
who arrive at location ${\bf 0}$ at a given time step must die at the next time
step. But by Polya's theorem {\sl all} random walkers visit location ${\bf 0}$
repeatedly. Hence, the total number of random walkers $N_t$ must vanish as $t\to
\infty$. On the other hand, if $D>2$ we have $\Pi_{\bf 0} <1$. Thus, the
fraction $1-\Pi_{\bf 0}$ of random walkers who originate in location ${\bf 0}$
{\sl never return} to ${\bf 0}$. These random walkers never die because $z=1$.
Hence, $N_t$ approaches a finite positive number as $t\to\infty$.

There is a change in the form of the transition at $D=4$. When $D<4$ the slope
of $B_2$ is continuous. However, when $D>4$ an elbow appears in $B_2$ at the
critical value $a_{\rm c}$. Specifically, when $D>4$ the slope of $B_2$ is $0$
for $0\leq a<a_{\rm c}$; just above $a_{\rm c}$ the slope abruptly becomes $-
\Pi_{\bf 0}/(T-1)$. Here, $T$ is the expected time for a random walker who
begins at location ${\bf 0}$ to return to ${\bf 0}$ when $a=1$ and $z=1$. A
quick heuristic argument yields this result \cite{rw2}: When $D<4$ we know that
$T$, the first temporal moment of $C_{{\bf 0},t;{\bf 0}}$, is divergent; $T$ is
finite when $D>4$. In a steady state all random walkers at location ${\bf 0}$
leave this location and in a $z=1$ model only the fraction $\Pi_{\bf 0}$ ever
return. The random walkers who return to location ${\bf 0}$ do so in $T$ steps
on average. These returning random walkers experience a death rate $z$ for $T-1$
of these $T$ steps. Thus, the expected number of random walkers who actually
return to location ${\bf 0}$ is decreased by the factor $z^{T-1}$. Hence, after
$T$ steps we expect to find the number of walkers at location ${\bf 0}$ to be
multiplied by the factor $a\Pi_{\bf 0} z^{T-1}$. The condition that there be a
steady state is therefore $a\Pi_{\bf 0}z^{T-1}=1$. Near the critical point $a=a_
{\rm c}+\delta$ and $z=1-\epsilon$ as $\delta,\epsilon\to 0^+$. To first order
in $\delta$ and $\epsilon$ the steady-state condition gives the slope of $B_2$
as $a\to a_{\rm c}^+$. Note that this argument is valid only for $a\geq a_{\rm
c}$.

While the above argument is only valid in the neighborhood of $a_{\rm c}$, the
same reasoning yields the entire curve $z(a)$ in the limit $D\to\infty$. In this
limit $T=2$. Hence, the steady-state condition gives $z=a_{\rm c}/a$.

To determine the nature of the phase transition in $F(a)$ along $B_2$, we let
$g_{\bf n}=\lim_{t\to\infty} G_{{\bf n},t}$. The steady-state distribution
$g_{\bf n}$ satisfies the difference equation
\begin{eqnarray}
g_{\bf n}=\cases{
z\sum_{\{\sigma \}_{\bf n}} P(\sigma\to {\bf n}) g_{\sigma}& $({\bf
n}\not \in \{\sigma\}_{\bf 0}),$\cr 
\noalign{\smallskip}
a P({\bf 0}\to{\bf n}) g_{\bf 0}+
z\sum_{\{\sigma\neq{\bf 0}\}_{\bf n}} P(\sigma\to {\bf n})
g_{\sigma}& $({\bf n}\in\{\sigma\}_{\bf 0}),$\cr }
\label{e11}
\end{eqnarray}
which is the time-independent version of Eq.~(\ref{e4}). For a steady-state
solution having a finite number of random walkers the sum $\sum_{\bf n}g_{\bf n}
$ exists. If we sum both sides of Eq.~(\ref{e11}) over all locations, we obtain
\begin{eqnarray}
\sum_{\bf n} g_{\bf n}= (a-z)g_{\bf 0} + z\sum_{\bf n} g_{\bf n}.
\label{e4.2}
\end{eqnarray}
Since the sum $\sum_{\bf n} g_{\bf n}$ is nonzero we obtain a formula for $F(a)$
in terms of $a$ and $z(a)$:
\begin{eqnarray}
{}F(a)={g_{\bf 0}\over \sum_{\bf n} g_{\bf n}}= {1-z(a)\over a-z(a)}.
\label{e4.3}
\end{eqnarray}
The result in Eq.~(\ref{e4.3}) is valid on the curved part of $B_2$; it is also
valid on the straight-line portion of $B_2$ when $D>2$. On $B_2$ we must treat
$z$ as a function of $a$. We emphasize this dependence by writing $z(a)$ and by
treating the fraction $F$ as a function of $a$ only.

We have studied Eq.~(\ref{e11}) for several types of lattices. We have
determined that along the curve $B_2$ the steady-state distribution fraction
$F(a)$ behaves like
\begin{eqnarray}
{}F(a)\sim {\rm C}(D)(a-a_{\rm c})^{\nu}\quad (a\to a_{\rm c}^+,~D\neq 2,4).
\label{e12}
\end{eqnarray}
The multiplicative constant ${\rm C}(D)$ depends on the specific choice of
lattice. However, the critical exponent $\nu$ is universal and only depends on
the dimension $D$. Using the probabilities in Eqs.~(\ref{e22}) for the
hyperspherical lattice, we derive in Ref.~\cite{rw2}
\begin{eqnarray}
\nu=\cases{{D\over 2-D}&$(0<D<2),$\cr
\noalign{\smallskip}
{2\over D-2}&$(2<D<4),$\cr 
\noalign{\smallskip}
1 &$(D>4).$\cr}
\label{e13}
\end{eqnarray}
We have verified the universality of these results numerically for more
complicated lattices. We have computed the critical exponent $\nu$ for three
types of random walks that are not analytically tractable: (i) a spherically
symmetric random walk using the probabilities in Eqs.~(\ref{e20}) for various
values of $D$; (ii) a conventional random walk on a $D=3$ cubic lattice; (iii)
a random walk on a recursively defined fractal lattice (Sierpinski carpet)
\cite{Feder} with dimension $D={\ln 12\over\ln 4}$. The agreement between the
predicted value of the critical exponent $\nu$ in Eq.~(\ref{e13}) and that
obtained by computer simulation is excellent, which provides strong evidence
that the second-order phase transition at $a_{\rm c}$ is universal.

There is no critical exponent for the special cases $D=2,4$; for the case of a
spherically symmetric random walk using the probabilities in Eqs.~(\ref{e22}) we
find that \cite{rw2} as $a\to a_{\rm c}^+$
\begin{eqnarray}
{}F(a)\sim\cases{\displaystyle
{{\rm constant}\over a-a_{\rm c}} e^{-{2\over a-a_{\rm c}}}&$(D=2),$\cr 
\noalign{\smallskip}
\displaystyle {a-a_{\rm c} \over 12\log \left({1\over a-a_{\rm
c}}\right )}  &$(D=4).$\cr}
\label{e14}
\end{eqnarray}

This work can be applied to the study of polymer growth in the vicinity of an
attractive boundary \cite{poly}. For a hyperspherical boundary of radius $m$,
this phenomenon is described by a partial difference equation of the same form
as that in Eq.~(\ref{e11}) \cite{BoMo}. Let $P(\kappa)$ represent the fraction
of a polymer that is adsorbed as a function of the attractive boundary potential
$\kappa$. This fraction is analogous to $F(a)$ in Eq.~(\ref{e4.3}). The 
potential $\kappa$ is a monotonic function of the birth rate $a$, and near the
critical point \cite{rw3}
\begin{eqnarray}
P(\kappa)\propto -{dz(a)\over da}\sim C(D) (a-a_{\rm c})^{\nu'}.
\label{new1}
\end{eqnarray}
We find that $\nu'=\nu$ ($D<2$) and $\nu'=\nu-1$ ($D>2$), where $\nu$ is given
in Eq.~(\ref{e13}). Thus, the polymer adsorption fraction $P(\kappa)$ exhibits
a first-order transition for $D>4$ and a tricritical point with logarithmic
scaling at $D=4$. This behavior for $P(\kappa)$ is illustrated in Fig.~\ref{f2}.

The analysis in this paper allows us to make a physical prediction for systems
where excluded-volume effects can be neglected: The adsorption fraction for a
solution of polymers growing near an approximately spherical attractive boundary
(such as a cell membrane) exhibits a second-order phase transition with linear
scaling.

Two of us, CMB and PNM, wish to thank the U.S. Department of Energy
for financial support under grant number DE-FG02-91-ER40628. SB
also thanks the U.S. Department of Energy for support under grant
number DE-AC02-76-CH00016.

\figure{
Generic phase diagram for the $(a,z)$ plane. Shown on the diagram are the
boundary curves $B_1$ and $B_2$. To the left of $B_1$ and on $B_1$ the fraction
of random walkers at location ${\bf 0}$, $F_t$, approaches $0$ as $t\to\infty$;
to the right of $B_1$ this fraction approaches a finite positive number as
$t\to\infty$. Above $B_2$ the total number of random walkers, $N_t$ diverges as
$t\to\infty$; below $B_2$ the total number of walkers approaches $0$ as
$t\to\infty$. On $B_2$ the distribution of random walkers approaches a steady
state as $t\to\infty$. The critical point $(a_{\rm c}, z_{\rm c}=1)$
lies at the intersection of $B_1$ and $B_2$. \label{f1}}

\figure{
Plot of the adsorption fraction $P(\kappa)$. For increasing $D<2$ the scaling 
exponent increases and the transition becomes weaker until at $D=2$ exponential
scaling is obtained. For increasing $D>2$ the scaling exponent decreases and the
transition becomes stronger again; this is compensated by an increase in the
critical binding potential $\kappa$ that is required to bring about the
transition. At $D=4$ we observe a tricritical point with logarithmic scaling.
{}For $D>4$ the transition is first order, indicated by a discontinuity
(green shaded region) in $P(\kappa)$ across the critical point. \label{f2}}
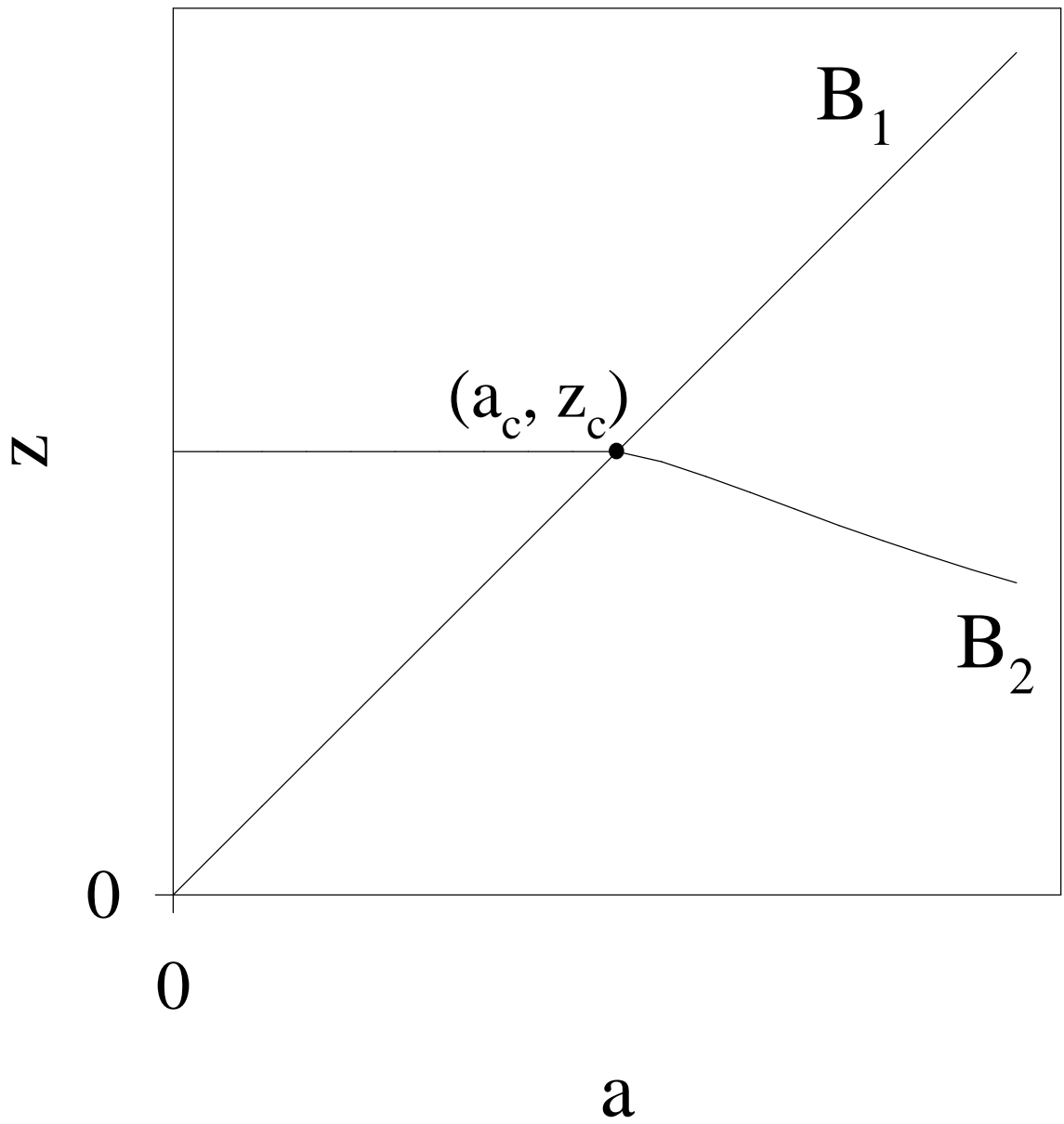
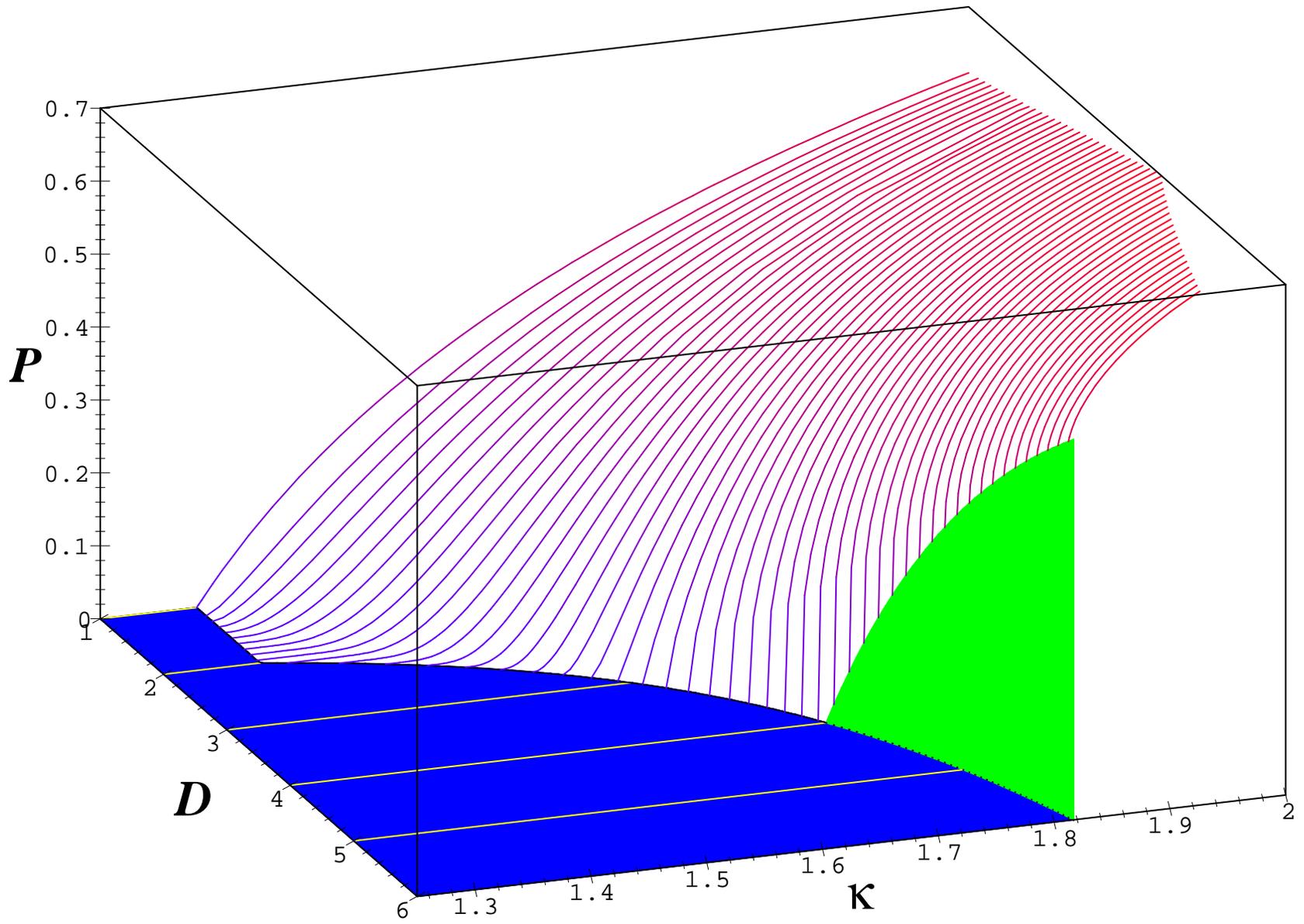
\end{document}